\documentclass[12pt]{article}
\usepackage{graphicx}
\begin{document}

\begin{center}

{\Large \bf Optical Activities as Computing Resources for Space-time
 Symmetries}

\vspace{6mm}
Y. S. Kim\footnote{electronic address: yskim@umd.edu} \\
Center for Fundamental Physics, University of Maryland, College Park,
Maryland 20742, U.S.A.

\end{center}

\vspace{6mm}

\begin{abstract}
  It is known that optical activities can perform rotations.  It
  is shown that the rotation, if modulated by attenuations, can
  perform symmetry operations of Wigner's little group which
  dictates the internal space-time symmetries of elementary
  particles.
\end{abstract}

\vspace{15mm}


\section{Introduction}\label{intro}

Polarization optics serves as analog computers for the geometry of
ellipsometry~\cite{azzam77} and the Poincar\'e sphere~\cite{bross98}.
They also can perform algebras of two-by-two and four-by-four matrices
known as Jones and Mueller matrices respectively~\cite{azzam77,theo79}.
It was established recently that these matrices correspond to the
two-by-two and four-by-four representations of the Lorentz group
which serves as the mathematical framework for Einstein's special
relativity~\cite{hkn97josa,hkn97,bgkn04}.

Optical activities are known to speak the language of rotations.  In
the real world, all optical rays go through attenuations.  If the
attenuation is axially symmetric, it does not raise additional
mathematical problems.

In this paper, let us assume that the optical ray propagates along
the $z$ direction, and that the polarization rotates on the $xy$
plane.  If the attenuation along the $x$ direction is different
from that along the $y$ direction, the combined effect of this
asymmetric attenuation and the rotation around the $z$ axis can
perform an interesting mathematical operation.

It is a simple matter to construct a rotation matrix for a given
value of propagation distance $Z$.  So is the matrix for the
asymmetric attenuation.  However, the problem becomes nontrivial
when these two effects are combined at a microscopic scale with a
small value of $z$, and this combined effect is repeated to make
up the finite value of $z$.

It is shown in this paper that the resulting mathematics not only
allows to make analytical calculations of the optical activities
with asymmetric attenuation effects, but also provides a
computational instrument for Wigner's little group which dictates
the internal space-time symmetries of elementary particles.

In 1939, Wigner noted a particle can have internal variables in
addition to its energy and momentum~\cite{wig39}.  For instance,
an election can have its spin degrees of freedom, in addition to
its momentum and energy.  Photons can have helicity and gauge
degrees of freedom.  Wigner formulated this symmetry problem by
introducing a three-parameter subgroup of the Lorentz group which
preserves the four-momentum of a given particle.  This subgroup
is called Wigner's little group in the literature.

For a massive particle, the little group is a Lorentz-boosted
rotation group.  For a massless particle, it is like (locally
isomorphic to) the two-dimensional Euclidean group.  For a
tachyon with a space-like four-momentum, the little group is
a Lorentz-boosted boost matrix, where the two boosts are made
along perpendicular directions~\cite{wig39,knp86}.

It is now possible to understand optical activities in terms
of Wigner's little group.  Conversely, the optical activity can
serve as an analog computer for internal space-time symmetries
of elementary particles.

In Sec.~\ref{formul}, we formulate the problem in terms of
two-by-two matrices applicable to the Jones vector.  It is not
difficult to write matrices performing rotations and attenuations
separately.
In Sec.~\ref{compu}, we compute the transformation matrix if those
two operations are performed at a microscopic scale, and are
accumulated to a macroscopic scale.
In Sec.~\ref{little}, it is shown that the transformation matrices
correspond to those of Wigner's little group which dictates
internal space-time symmetries of elementary particles.

\section{Formulation of the Problem}\label{formul}

Let us start with a light wave taking the form
\begin{equation}\label{expo1}
\pmatrix{E_{x} \cr E_{y}} =
\pmatrix{A \exp{\left\{i(kz - \omega t + \phi_{1})\right\}}  \cr
B \exp{\left\{i(kz - \omega t + \phi_{2})\right\}}} .
\end{equation}
This ray can go through rotations around the $z$ axis.  It can also
go through $xy$ asymmetric phase shifts and attenuations.  The
mathematics of these aspects is known as the Jones matrix formalism.
If there are decoherence effects between the $x$ and $y$ components,
the mathematics can be extended to the four-by-four Mueller matrix
formalism~\cite{bross98}.

It has been recently established that the Jones and Muller formalisms
constitute the two-by-two and four-by-four representations of the
Lorentz group~\cite{hkn97josa, hkn97}.  In the two-by-two case, the
transformation matrix is generated by three Pauli matrices
\begin{equation}\label{rot11}
\sigma_1 = \pmatrix{0 & 1 \cr 1 & 0}, \qquad
\sigma_2 = \pmatrix{0 & -i \cr i & 0}, \qquad
\sigma_3 = \pmatrix{1 & 0 \cr 0 & -1} ,
\end{equation}
plus three squeeze generators
\begin{equation}\label{sq00}
\tau_1 = i\sigma_1 = \pmatrix{0 & i \cr i & 0}, \quad
\tau_2 = i\sigma_2 = \pmatrix{0 & 1 \cr -1 & 0}, \quad
\tau_3 = i\sigma_3 = \pmatrix{i & 0 \cr 0 & -i} .
\end{equation}
There are therefore six generators.  They form a closed set of
commutation relations.  In mathematical terms, they form the
Lie algebra for the SL(2,c) group which is locally isomorphic
to the six-parameter Lorentz group.

The three matrices of Eq.(\ref{rot11}) generate the rotation
subgroup of the Lorentz group.  For possible computer
applications, we are interested in the subgroup which produces
real transformation matrices.  Imaginary numbers are not
convenient for computer mathematics.

Among the above six generators given in Eqs.(\ref{rot11}) and
(\ref{sq00}),  $\sigma_2, \tau_3$, and $\tau_1$ are pure imaginary
and can generate real transformation matrices.  They satisfy
the following closed set of commutation relations.
\begin{equation}
\left[\sigma_2, \tau_3\right] = 2i \tau_1, \quad
\left[\tau_3, \tau_1\right] = -2i \sigma_2, \quad
\left[\tau_1, \sigma_2\right] = 2i \tau_3.
\end{equation}
This group generated by these three matrices is called the
$Sp(2)$ group and is applicable to many optical instruments and
optical processes, either directly or indirectly through its
isomorphism with the $S(1,1)$ group for squeezed states.

Within this framework, we are dealing with rotation matrices
of the form
\begin{equation}\label{theta}
R(\theta) = \pmatrix{\cos\alpha & -\sin\alpha \cr
                          \sin\alpha & \cos\alpha} ,
\end{equation}
and squeeze matrices of the form
\begin{equation}\label{beta}
S(\alpha) = \pmatrix{\exp{(\beta)} & 0 \cr 0 & \exp{(-\beta)}} ,
\end{equation}
and their multiplications.  All the matrices in this
representation are real.

Indeed, optical activities can be described by these real
matrices.  The polarization goes through the rotation
\begin{equation}\label{rot22}
R(\gamma z) = \pmatrix{\cos(\gamma z) & -\sin(\gamma z) \cr
  \sin(\gamma z) & \cos(\gamma z)} ,
\end{equation}
as the ray propagates along the z direction.  This matrix is
applicable to the $x$ and $y$ components of the polarization,
and the rotation angle increases as $z$ increases.

The optical ray is expected to be attenuated due to absorption
by the medium.  The attenuation coefficient in one transverse
direction could be different from the coefficient along the
other direction.  Thus, if the rate of attenuation along the
$x$ direction is different from that along $y$ axis, this
asymmetric attenuation can be described by
\begin{equation}\label{atten}
\pmatrix{\exp{\left(-\mu_{1}z \right)} & 0 \cr
    0 & \exp{\left(-\mu_{2}z \right)}}
 = e^{-\lambda z} \pmatrix{\exp{(\mu z)} & 0 \cr
                        0 & \exp{(-\mu z)}} ,
\end{equation}
with
\begin{eqnarray}
&{}& \lambda = \frac{\mu_{2} + \mu_{1}}{2} , \nonumber \\[1ex]
&{}& \mu = \frac{\mu_{2} - \mu_{1}}{2} .
\end{eqnarray}
The exponential factor $\exp{(-\lambda z)}$ is for the overall
attenuation, and the matrix
\begin{equation}
  \pmatrix{\exp{(\mu z)} & 0 \cr 0 & \exp{(-\mu z)}} ,
\end{equation}
performs a squeeze transformation.  This matrix expands the $x$
component of the polarization, while contracting the $y$ component.
We shall call this the squeeze along the $x$ direction.

The squeeze does not have to be along the $x$ direction.  It can
be in the direction which makes an angle $\theta$ with the
$x$ axis.  The squeeze matrix then becomes
\begin{equation}
S(\theta, \mu z) = \pmatrix{\cos\theta & -\sin\theta \cr
        \sin\theta & \cos\theta}
        \pmatrix{\exp{(\mu z)} & 0 \cr 0 & \exp{(-\mu z)}}
\pmatrix{\cos\theta & \sin\theta \cr -\sin\theta & \cos\theta} ,
\end{equation}
which can be compressed to one matrix
\begin{equation}
 \pmatrix{\cosh(\mu z) + \sin(2\theta) \sinh(\mu z) &
  \sin(2\theta) \sinh(\mu z) \cr
  \sin(2\theta) \sinh(\mu z) &
  \cosh(\mu z) - \sin(2\theta)\sinh(\mu z)} .
\end{equation}
If $\theta = 45^o$, this matrix becomes
\begin{equation}\label{sq11}
S(\pi/4, \mu z) = \pmatrix{\cosh(\mu z) & \sinh(\mu z) \cr
  \sinh(\mu z) & \cosh(\mu z) } .
\end{equation}
We shall work with this form of squeeze matrix in the following
discussion, and use the notation $S(\mu z)$ without angle for the
above expression.  Thus, if the squeeze is made along the $x$
axis, the squeeze matrix is
\begin{equation}
S(0, \mu z) = R(-\pi/4, \mu z) S(\mu z) R(\pi/4, \mu z) .
\end{equation}

If this squeeze is followed the rotation of Eq.(\ref{rot22}),
the net effect is
\begin{equation}
e^{-\lambda z} \pmatrix{\cos(\gamma z) & -\sin(\gamma z) \cr
  \sin(\gamma z) & \cos(\gamma z)}
\pmatrix{\cosh(\mu z) & \sinh(\mu z) \cr
  \sinh(\mu z) & \cosh(\mu z) } ,
\end{equation}
where $z$ is in a macroscopic scale, perhaps measured in
centimeters.  However, this is not an accurate description
of the optical process.

This happens in a microscopic scale of $z/N$, and becomes
accumulated into the macroscopic scale of $z$ after the
$N$ repetitions, where $N$ is a very large number.  We are
thus led to the transformation matrix of the form
\begin{equation}\label{trans}
M(\gamma,\mu,z)= \left[e^{-\lambda z/N}S(\mu z/N)
                   R(\gamma z/N)\right]^N .
\end{equation}
In the limit of large $N$, this quantity becomes
\begin{equation}
e^{-\lambda z/} \left[\pmatrix{1 & \mu z/N \cr \mu z/N & 0}
\pmatrix{1 & - \gamma z/N \cr \gamma z/N & 1}\right]^N .
\end{equation}
Since $\gamma z/N$ and $\mu z/N$ are very small,
\begin{equation}\label{mat11}
M(\gamma,\mu,z)= e^{-\lambda z} \left[\pmatrix{1 & 0 \cr 0 & 1}
   + \pmatrix{0 & -(\gamma - \mu) \cr
  (\gamma + \mu) & 0}\frac{z}{N} \right]^N .
\end{equation}
For large $N$, we can write this matrix as
\begin{equation}\label{m33}
M(\gamma,\mu,z) = e^{-\lambda z} \exp{\left( G z \right)} ,
\end{equation}
with
\begin{equation}\label{g11}
G = \pmatrix{0 & -(\gamma - \mu) \cr (\gamma + \mu) & 0} .
\end{equation}

The remaining problem is to calculate the exponential form
$\exp{\left( G z \right)}$ by making a Taylor expansion.  We
thus need to compute $G^N$.  This is a trivial problem if $G$ is
diagonal or can be diagonalized by a similarity transformation
of a diagonal matrix.  The problem arises because this is not
always the case.

\section{Computation of the Transformation Matrix}\label{compu}

We are interested in computing the exponential form of
Eq.(\ref{mat11}).  If $\gamma$ in Eq.(\ref{g11}) is greater
than $\mu$, the off-diagonal elements have opposite signs, and
we can write $G$ as
\begin{equation}\label{g22}
  G = k \pmatrix{0 & -\exp{(2\eta)} \cr \exp{(-2\eta)}  & 0} ,
\end{equation}
with
\begin{eqnarray}\label{gammak}
&{}&  k = \sqrt{\gamma^2 - \mu^2}, \nonumber \\[1ex]
&{}&  \exp{(2\eta)} = \sqrt{\frac{\gamma + \mu}{\gamma - \mu}},
\end{eqnarray}
or conversely
\begin{equation}\label{gamma}
\gamma = k \cosh(2\eta), \qquad \mu = k \sinh(2\eta) .
\end{equation}

If $\mu$ is greater than $\gamma$, the off-diagonal elements have
the same sign.  We can then write $G$ as
\begin{equation}\label{g33}
 G = k \pmatrix{0 & \exp{(2\eta)} \cr \exp{(-2\eta)}  & 0} ,
\end{equation}
with
\begin{eqnarray}\label{muk}
&{}&  k = \sqrt{\mu^2 - \gamma^2}, \nonumber\\[1ex]
&{}& \exp{(2\eta)} = \sqrt{\frac{\mu + \gamma}{\mu - \gamma}} ,
\end{eqnarray}
or
\begin{equation}
\gamma = k\sinh(2\eta), \qquad \mu = k\cosh(2\eta) .
\end{equation}

If $\gamma = mu$, the upper-right element of the $G$ matrix has
to vanish, and it becomes
\begin{equation}
\pmatrix{0 & 0 \cr 2\gamma  & 0} .
\end{equation}
As $\mu$ becomes larger from $\mu < \gamma$ to $\mu > \gamma$, the
$G$ matrix has to go through this triangular form.

We are now ready to compute the exponential form
\begin{equation}\label{expo}
\exp{\left(z G\right)} .
\end{equation}
The problem is whether it is possible to obtain an analytical
expression of the above quantity.  The usual procedure is to write
a Taylor expansion.  For this purpose, we need to calculate $G^N$.
We can manage this calculation when $N = 2$.  However, for an
arbitrary large integer $N$, it is not a trivial problem.  This
is exactly the problem we would like to address in this section.

If $\gamma$ is greater than $\mu$, we write $G$ of Eq.(\ref{g22})
as
\begin{equation}
G = k \pmatrix{e^{\eta} & 0 \cr 0 & e^{-\eta}}
 \pmatrix{0 & -1 \cr 1 & 0}
 \pmatrix{e^{-\eta} & 0 \cr 0 & e^{\eta}} ,
\end{equation}
with $\eta$ given in Eq.(\ref{gammak}).  This is a similarity
transformation of
\begin{equation}
\pmatrix{0 & -1 \cr 1 & 0} ,
\end{equation}
with respect to a squeeze matrix
\begin{equation}\label{sq33}
B(\eta) = \pmatrix{\exp{(\eta)} & 0 \cr 0 & \exp{(-\eta)}} .
\end{equation}
The role of this squeeze matrix is quite different from that of
Eq.(\ref{sq11}).  It does not depend on $z$.

Let us go back to the $G$ matrix, We can write $G^N$ as
\begin{equation}
 k^N \pmatrix{e^{\eta} & 0 \cr 0 & e^{-\eta}}
    \pmatrix{0 & -1 \cr 1 & 0}^N
 \pmatrix{e^{-\eta} & 0 \cr 0 & e^{\eta}} ,
\end{equation}
and
\begin{equation}
\exp{\left[k\pmatrix{0 & -1 \cr 1 & 0}\right]} =
 \pmatrix{\cos(kz) & -\sin(kz) \cr \sin(kz) & \cos(kz)} .
\end{equation}
Thus, the exponential form $\exp{(Gz)}$ in of Eq.(\ref{expo})
becomes
\begin{equation}
 \pmatrix{e^{\eta} & 0 \cr 0 & e^{-\eta}}
 \pmatrix{\cos(kz) & -\sin(kz) \cr \sin(kz)  & \cos(kz)}
 \pmatrix{e^{-\eta} & 0 \cr 0 &  e^{\eta}} ,
\end{equation}
and the transformation matrix of Eq.(\ref{trans}) takes the form
\begin{equation}
M(\gamma,\mu,z) = e^{-\lambda z}
           \pmatrix{\cos(kz) & - e^{2\eta} \sin(kz) \cr
           e^{-2\eta} \sin(kz)  & \cos(kz)} ,
\end{equation}
with $k$ and $\eta$ given in Eq.(\ref{gammak}).

If $\mu$ is greater than $\gamma$, the off-diagonal elements of
Eq.(\ref{g11}) have the same sign, but we can go through a
similar calculation.  The result is
\begin{equation}
M(\gamma,\mu,z) = e^{-\lambda z}
           \pmatrix{\cosh(kz) &  e^{2\eta} \sinh(kz) \cr
           e^{-2\eta} \sinh(kz)  & \cosh(kz)} ,
\end{equation}
with $k$ and $\eta$ given in Eq.(\ref{muk}).

If $\gamma$ and $\mu$ are equal, the $G$ matrix becomes
\begin{equation}\label{tri22}
  G = \pmatrix{0 & 0 \cr 2\gamma & 0} ,
\end{equation}
with the property
\begin{equation}
  G^2 = \pmatrix{0 & 0 \cr 2\gamma & 0}^2 = 0 ,
\end{equation}
and the transformation matrix becomes
\begin{equation}
    \pmatrix{1 & 0 \cr 2\gamma z & 1} .
\end{equation}

Let us go back to the case with $\gamma > \mu$.   We can then
gradually increase the parameter $\mu$ to a value greater than
$\gamma$, which means from Eq.(\ref{gammak}) to Eq.(\ref{muk}).
This involves a singularity in the expression of $\exp{(2\eta)}$
in these equations.  This is a complicated mathematical
issue~\cite{gk03}, but we can avoid the problem using the
variables $\mu$ and $\gamma$.

\section{Space-time Symmetries spoken by Optical
              Activities}\label{little}

As was mentioned in Sec.(\ref{intro}), the Lorentz group provides
the basic mathematical framework for polarization optics.  The
Lorentz group was used earlier by Einstein to formulate his
special theory of relativity.  In 1905, Einstein considered only
point particles.  After the formulation of quantum mechanics in
1927, it was found that particles can have internal space-time
structures.

If a given particle has internal space-time symmetries, such as
electron spin and quark distribution inside a hadron, we have
to rely on Wigner's little groups~\cite{wig39}.  If the particle
is massive, there is a Lorentz frame in which the particle is at
rest.  In this frame, the four-momentum remains invariant under
rotations.  However, its spin can change its orientation.
Wigner's little group in this case is like (locally isomorphic to)
the three-dimensional rotation group.  We call this the $O(3)$-like
little group for massive particles.

We do not observe particles with space space-like momentum or
moving faster than light, but they play important roles in
physical theories. We need those space-like particles in Feynman
diagrams.  For a particle of this type, there is the Lorentz
frame where the momentum does not have its time-like component.
It has its space component along a given direction.  This
four-momentum is invariant under Lorentz boots along the two
perpendicular directions.  The subgroup in this case is the
Lorentz group applicable to two space coordinates and one time
variable.  We call this $O(2,1)$-like subgroup.

If the particle is massless, like photons, there are no Lorentz
frames in which it is rest or with zero time-like component.
For this case, Wigner in 1939 observed that there is a
three-parameter subgroup of the Lorentz group which leaves the
four-momentum invariant, and that its algebraic property is the
same as that of the two-dimensional Euclidean group.  We call
this $E(2)$-like little group for massless particles.

Let us now translate the two-by-two matrices given in
Secs.~\ref{formul} and \ref{compu} into the language of
four-by-four Lorentz transformation matrices applicable to the
Minkowski space of $(z,y,z,t)$.  In this convention, the
momentum-energy four-vector is $\left(p_x, p_y, p_z, E\right).$
If the particle moves along the $z$ direction, this four-vector
becomes
\begin{equation}\label{mom11}
\left(0, 0, p, \sqrt{p^2 + m^2}\right) ,
\end{equation}
in the unit system where $c = 1,$ where $m$ is the particle mass.  We
can obtain this four-vector by boosting a particle at rest with the
four-momentum
\begin{equation}\label{mom22}
(0, 0, 0, m) ,
\end{equation}
with the boost matrix
\begin{equation}\label{b11}
 B(\eta) = \pmatrix{1 & 0 & 0 & 0 \cr
   0 & 1 & 0 & 0   \cr
   0 & 0 & \cosh(2\eta) & \sinh(2\eta) \cr
        0 & 0 & \sinh(2\eta) & \cosh(2\eta)} ,
\end{equation}
where
\begin{equation}
\tanh(2\eta) = \frac{p}{\sqrt{p^2 + m^2}} .
\end{equation}
It is known that this boost matrix corresponds to the squeeze matrix
$B(\eta)$ of Eq.(\ref{sq33})\cite{hkn97}.

Now the four-momentum of Eq.(\ref{mom22}) is invariant under the
rotation matrix
\begin{equation}\label{rot55}
 R(\gamma z) = \pmatrix{\cos(2\gamma z) & 0 & \sin(2\gamma z) & 0
   \cr 0 & 1 & 0 & 0 \cr
    -\sin(2\gamma z) & 0 & \cos(2\gamma z) & 0 \cr 0 & 0 & 0 & 1} .
\end{equation}
Thus, the matrix
\begin{equation}\label{wlg11}
B(\eta) R(\gamma z) B(-\eta)
\end{equation}
leaves the four-momentum of Eq.(\ref{mom11}) invariant.
While this matrix performs a rotation around the $y$ axis in the
particle's rest frame, we can also rotate this four-momentum
around the $z$ axis without changing it.  This is what Wigner's
little group is about for the particle with mass $m$.

Although the matrix of Eq.(\ref{wlg11}) does not change the
momentum, it rotates the spin direction of the particle in its
rest frame.  This is why the little group is not a trivial
mathematical device.

It is known that the rotation matrix of Eq.(\ref{rot55}) corresponds
to the rotation matrix of Eq.(\ref{rot22})~\cite{hkn97}.  Thus
the two-by-two rotation matrix of Eq.(\ref{rot22}), together with the
squeeze matrix of Eq.(\ref{sq33}), generates the little group for
particles with non-zero mass.

If the particle has a space-like momentum, we can start with the
four-momentum
\begin{equation}
  (0, 0, p, E) ,
\end{equation}
where $E$ is smaller than $p$, which it can be brought to the
Lorentz frame where the four-vector becomes
\begin{equation}\label{mom33}
  (0, 0, p, 0) .
\end{equation}
The boost matrix takes the same form as Eq.(\ref{b11}), with
\begin{equation}
\tanh(2\eta) = \frac{E}{p} .
\end{equation}
The four-momentum of Eq.(\ref{mom33}) is invariant under the boost
\begin{equation}
 S(\mu z) = \pmatrix{\cosh(2\mu z) & 0 & 0 & \sinh(2\mu z)  \cr
   0 & 1 & 0 & 0   \cr   0 & 0 & 1 & 0 \cr
   \sinh(2\mu z) & 0 & 0 & \cosh(2\mu z) }
\end{equation}
along the $x$ direction.  Here again the four-momentum of
Eq.(\ref{mom33}) is invariant under rotations around the $z$
axis.

The above four-by-four matrix corresponds to the two-by-two squeeze
matrix of Eq.(\ref{sq11}) applicable to optical activities~\cite{hkn97}.
Thus, this squeeze matrix, together with the squeeze matrix of
Eq.(\ref{sq33}), generate the little group for particles with space-like
momentum.

Let us finally consider a massless particle with it four-momentum
\begin{equation}
(0, 0, p, p) .
\end{equation}
It is invariant under the rotation around the $z$ axis.  In addition, it
is invariant under the transformation
\begin{equation}\
\pmatrix{1 & 0 & -2\gamma & 2\gamma \cr 0 & 1 & 0 & 0 \cr
               2\gamma & 0 &  1 -2\gamma^2 & 2\gamma^2 \cr
               2\gamma & 0 &   -2\gamma^2 & 1 + 2\gamma^2 }.
\end{equation}
This four-by-four matrix has a stormy history~\cite{knp86,kiwi90jm},
but the bottom line is that it corresponds to the triangular matrix
of Eq.(\ref{tri22}), and the variable $\gamma$ performs gauge
transformations.

It is interesting to note that optical activities can act as
computational devices for the internal space-time symmetries of
elementary particles.

\section*{Concluding Remarks}

Each human being has ten fingers.  This is the origin of our
decimal system.  Vacuum tubes can do binary logic, and this is
how electronic computer was developed.  Quantum two-level
systems can do more than the vacuum tube can.  This is why we
are interested in quantum computers these days.  Indeed,
computers are based on the mathematical language spoken by
nature.

Traditionally, polarization optics is known to produce the
geometry of ellipse and that of the Poincar\'e sphere.  It
also produces the algebra of two-by-two and four-by-four
matrices.  In this paper, we started with rotations combined
with asymmetric attenuations in optical activities.  It was
shown in this paper that the optical activity speaks the
mathematical language of Wigner's little group dictating
internal space-time symmetries of elementary particles.

We have used in this paper some mathematical methods not
commonly seen in the conventional literature.   In calculating
the exponential form of a matrix, the usual procedure is to
diagonalize the matrix by a unitary transformation.  Then it
is possible to write a Taylor expansion of the diagonal matrix.

What should we do if the matrix cannot be diagonalized by a
unitary transformation?  Let us go back to the $G$ matrix
of Eq.(\ref{g11}).  If $\gamma > \mu$, it was possible to
bring the $G$  the form
\begin{equation}\label{m66}
k \pmatrix{0 & -1 \cr 1 & 0} ,
\end{equation}
where the similarity transformation matrix of Eq.(\ref{sq33}) is not
unitary.  It is a symmetric squeeze matrix.  In addition, we used
the property
\begin{equation}
 \pmatrix{0 & -1 \cr 1 & 0}^2 = \pmatrix{1 & 0 \cr 0 & 1}, \qquad
 \pmatrix{0 & -1 \cr 1 & 0}^3 = \pmatrix{0 & -1 \cr 1 & 0},
\end{equation}
to deal with the Taylor expansion.  For $\mu > \gamma$, we used
\begin{equation}\label{m77}
 \pmatrix{0 & 1 \cr 1 & 0}^2 = \pmatrix{1 & 0 \cr 0 & 1}, \qquad
 \pmatrix{0 & 1 \cr 1 & 0}^3 = \pmatrix{0 & 1 \cr 1 & 0},
\end{equation}

If $\mu = \gamma$, the $G$ matrix becomes triangular, and
\begin{equation}\label{m88}
\pmatrix{0 & 0 \cr 2\gamma & 0}^2 = 0.
\end{equation}
The Taylor expansion truncates.

Using these properties of two-by-two matrices, we were able to deal
with the problem even though not all of them can be diagonalized.
The triangular matrix of Eq.(\ref{m88}) is triangular and cannot be
diagonalized.  The matrix of Eq.(\ref{m77}) can be diagonalized with
the diagonal elements of $1$ and $-1$.  The two-by-two matrix of
Eq.(\ref{m66}) can also be diagonalized, but the eigenvalues are
the imaginary numbers $i$ and $-i$.  However, the imaginary numbers
are not too convenient for computer mathematics.  Thus, we had to
resort to the method presented in this paper.

\end{document}